\documentclass[twocolumn,preprintnumbers,amsmath,amssymb]{revtex4}
\usepackage{graphicx}
\usepackage{dcolumn}
\usepackage{bm}
\usepackage[usenames]{color}
\usepackage{color}
\def\>{\rangle}\def\<{\langle}
\def\togli#1{}
\def\comment#1{}

\begin{document}
\title{Bell measurements as a witness of a dualism in entanglement}
\author{E. Moreva$^{1,2}$, G. Brida$^1$, M. Gramegna$^1$, S. Bose$^3$, D. Home$^4$,  M. Genovese$^{1*}$}
\affiliation{\vbox{$^1$INRIM, strada delle Cacce 91, 10135 Torino,Italy} \vbox{$^2$International Laser Center of M.V.Lomonosov Moscow State University, 119991, Moscow, Russia}\vbox{$^3$Department of Physics and Astronomy, Univestity College London, Gower Street, London WCIE 6BT, United Kingdom}
\vbox{$^4$CAPSS, Physics Department, Bose Institute, Salt Lake, Sector V, Kolkata 700097, India}
\vbox{$^{*}$corresponding author: m.genovese@inrim.it}}
\begin{abstract}

We show how a property of dualism, which can exist in the entanglement of identical particles, can be tested in the usual photonic Bell measurement apparatus with minor modifications. Two different sets of coincidence measurements on the same experimental setup consisting of a Hong-Ou-Mandel interferometer demonstrate how the same two-photon state can emerge entanglement in the polarization or the momentum degree of freedom depending on the dynamical variables used for labeling the particles. Our experiment demonstrates how the same source can be used as both a polarization entangled state, as well as a dichotomic momentum entangled state shared between distant users Alice and Bob in accordance to which sets of detectors they access. When the particles become distinguishable by letting the information about one of the variables to be imprinted in yet another (possibly inaccessible) system or degree of freedom, the feature of dualism is expected to vanish. We verify this feature by polarization decoherence (polarization information in environment) or arrival time difference, which both respectively destroy one of the dual forms of entanglement.
\end{abstract}
\maketitle

\section{Introduction}
Quantum sources of entangled particles represent a fundamental component for different applications \cite{genovese} such as quantum computing \cite{cq1,cq2,paladino}, quantum communication based protocols such as dense coding \cite{dense}, teleportation \cite{telp}, quantum cryptography \cite{ekert} and quantum metrology \cite{metrology1, metrology2, metrology3, metrology4}. Likewise, the quantum indistinguishability of identical particles has varied applications in information processing \cite{KLM, boson-samp, bose-home, omar-paunkovic} and is currently quite a topical issue \cite{Tichy}. It is quite natural to expect that the entangled states of identical particles will have curious features by virtue of their quantum indistinguishability.
One such property, dubbed ``entanglement dualism" \cite{bose} was introduced recently for indentical quantum particles. It characterizes the entanglement of two identical particles by different interchangeable variables. Recently, one observation of this property \cite{Duan} has been made  and a novel application called ``entanglement sorting"  has been formulated \cite{Agarwal} .

A pure bipartite state described by the wave function $|\Psi(x_{1},x_{2})\>$ with variables $x_{1},x_{2}$ of particles $1,2$ is defined to be entangled if
\begin{equation}
|\Psi(x_{1},x_{2})\>\ne |\phi(x_{1})\>|\varphi(x_{2})\>.
\label{eq:entangled}
\end{equation}
Variables $x_{1},x_{2}$ can be discrete, like polarization of the particle or spin projection, or continuous, like direction of momentum.
Thus, the maximally entangled polarization state of two identical photons, occupying only one of two spatial modes each $k$ or $-k$  can be written in the form
\begin{equation}
|\Psi\>=\tfrac1{\sqrt{2}}(|H\>_{k}|V\>_{-k}-|V\>_{k}|H\>_{-k})
\label{eq:state_pol}
\end{equation}
The propagation directions (and photon's momentum) are commeasurable with the photon's polarization, therefore the state $|\Psi\>$ can be equally rewritten into a form
\begin{equation}
|\Psi\>=\tfrac1{\sqrt{2}}(|-k\>_{H}|k\>_{V}-|k\>_{H}|-k\>_{V})
\label{eq:state_mom}
\end{equation}
where roles of polarization and momentum coordinate variables were rearranged. Equations (\ref{eq:state_pol}) and (\ref{eq:state_mom}) capture the property of dualism: we can use the momenta $k$ and $-k$ as the ``which particle" label and then the polarization of the two photons will be found to be entangled (\ref{eq:state_pol}). Alternatively, one can use the polarization of the two photons as the ``which particle" label, in which case, their momenta will be entangled (\ref{eq:state_mom}) \cite{bose}.
In contrast to ``hyper-entanglement" \cite{hyper}, in which into a quantum state more than one variable is simultaneously entangled, dualism emerges only from the interchangeability of different dynamical variables used for labeling the particles and has a property of complementarity in sense that one cannot observe entanglement on both variables at the same time.  Following the general schematic of the apparatus of Ref.\cite{bose} an experimental observation of the duality has been made in a photonic experiment \cite{Duan}. However, it has not yet been noticed that a different, and potentially much simpler, experimental setup can also test the same duality.

In this paper, we show how the entanglement duality can be tested by minimally modifying the usual Bell state measurement apparatus and demonstrate this scheme experimentally. For this purpose in our set-up we evaluate the entanglement using witness operator for the polarization and momentum degrees of freedom and show that only for undistinguishable particles the entanglement manifest its feature irrespective of variables used for labeling the particles. Moreover, an interesting behaviour of the entanglement of the states displaying the dualism has not yet been fully tested. This is the fact that even if some form of distinguishability or decoherence reduces the entanglement in one of the two degrees of freedom involved in the duality, the entanglement in the other degree of freedom may still be fully retained. This can be regarded as a type of ``robustness" of the entanglement in the states displaying the duality. It may have the practical use that if there is a decoherence (or eavesdropping) in one degree of freedom, the state may still retain some usefullness -- communicating parties Alice and Bob may then resort to using the undecohered degree of freedom for quantum communications. The previous experiment on dualism \cite{Duan} has tested only for the disappearence of the momentum entanglement due to a distinguishability of the photons, while the polarization entanglement was retained. Here we add to this by also showing that even when the polarization entanglement is destroyed due to decoherence, the momentum entanglement remains a useful resource. For the detection of entanglement we use a set of local measurements and calculate an entanglement-witness operator. Different from  quantum tomography, this method does not provide a full reconstruction of the quantum state but allows one, with a minimal number of local measurements, to check if the entanglement is really present.
The entanglement-witness operators that we will use are: for the polarization degree of freedom

\begin{equation}
\begin{array}{cc}
W_{p}=|+45_{1}\>|+45_{2}\>\<+45_{1}|\<+45_{2}|+\\|-45_{1}\>|-45_{2}\>\<-45_{1}|\<-45_{2}|-
\\(|R_{1}\>|L_{2}\>\<R_{1}|\<L_{2}|+|L_{1}\>|R_{2}\>\<L_{1}|\<R_{2}|),
\end{array}
\label{eq:W_{p}}
\end{equation}
where 1,2 are numbers of the photons and eigenstates $|x^{\pm}_{1,2}\>$, $|y^{\pm}_{1,2}\>$ are defined as
\begin{equation}
\begin{array}{ccc}
|\pm45_{1,2}\>=\tfrac1{\sqrt{2}}(|H_{1,2}\>\pm|V_{1,2}\>),\\|R_{1,2}\>=\tfrac1{\sqrt{2}}(|H_{1,2}\> + i|V_{1,2}\>),
\\|L_{1,2}\>=\tfrac1{\sqrt{2}}(|H_{1,2}\> - i|V_{1,2}\>),
\end{array}
\label{eq:Wpol}
\end{equation}
and for the momentum degree of freedom
\begin{equation}
\begin{array}{cc}
W_{m}=|K^{+}_{1}\>|K^{+}_{2}\>\<K^{+}_{1}|\<K^{+}_{2}|+|K^{-}_{1}\>|K^{-}_{2}\>\<K^{-}_{1}|\<K^{-}_{2}|- \\
-(|K^{+}_{1}\>|K^{-}_{2}\>\<K^{+}_{1}|\<K^{-}_{2}|+|K^{-}_{1}\>|K^{+}_{2}\>\<K^{-}_{1}|\<K^{+}_{2}|),
\end{array}
\label{eq:W_{p}}
\end{equation}
where
\begin{equation}
|K^{\pm}_{1,2}\>=\tfrac1{\sqrt{2}}(|k_{1,2}\>\pm i|-k_{1,2}\>).
\label{eq:Wmom}
\end{equation}
To see the above witness operators in a slightly different notation, one can define the Pauli operator  $\sigma_x = |H\rangle \langle V| + |V\rangle \langle H|$, $\sigma_y = i(|V\rangle \langle H| - |H\rangle \langle V|)$, $\sigma_z = |H\rangle \langle H| - |V\rangle \langle V|$ for the polarization degree of freedom. Then the witness operator $W_p = 0.5(\sigma_x^1 \sigma_x^2 + \sigma_y^1 \sigma_y^2)$, where $1,2$ are numbers designating the photons. Analogously, by defining $\sigma_x = |K_1\rangle \langle K_2| + |K_2\rangle \langle K_1|$, $\sigma_y = i(|K_2\rangle \langle K_1| - |K_2\rangle \langle K_1|)$ for momentum degrees of freedom, we have $W_m=0.5(\sigma_x^1 \sigma_x^2 + \sigma_y^1 \sigma_y^2)$. Defining the value of entanglement/momentum witness like a mean of the modulus of expectation value of the witness operator it is easy to verify that $|<W_{p,m}>|> 0.5$ if polarization or momentum degrees of freedom are entangled, while for separable states $|<W_{p,m}>|\leq0.5$.

\section{The setup}

Let us consider a pure entangled state of two photons, created via spontaneous parametric down-conversion in nonlinear crystal with phase matching conditions of type II. If we consider the entangled state with polarization as the entangled variable, and momentum as the ``which-particle" label,  then such state can be written as (\ref{eq:state_pol}), while a state with momentum as the entangled variable and polarization as the indexing has a form (\ref{eq:state_mom}).
The dualism in entanglement can be expressed as:

\begin{equation}
\tfrac1{\sqrt{2}}(|H\>_{k}|V\>_{-k}-|V\>_{k}|H\>_{-k})=\tfrac1{\sqrt{2}}(|-k\>_{H}|k\>_{V}-|k\>_{H}|-k\>_{V})
\label{eq:dualism}
\end{equation}

\begin{figure}[ht!]
\begin{center}
\includegraphics[width=0.5\textwidth]{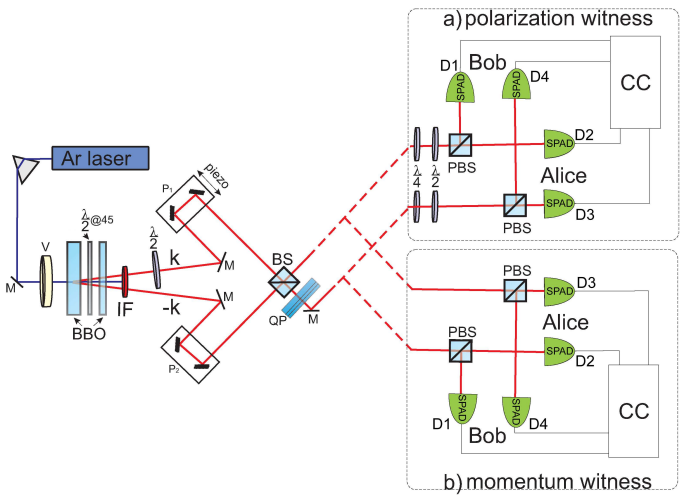}
\caption{Experimental setup for the measurement of the entanglement dualism. Blocks a and b serve for measuring the polarization witness and momentum witness correspondingly. Experimental scheme allows one to switch between two types of measurements. In scheme - Ar laser: argon  cw laser with wavelength 351 nm, M: mirror, V: vertical oriented Glan-Thompson prism, BBO: nonlinear barium borate crystals, IF: interfilter, $P_{1,2}$ optical trombones $\lambda/2$, $\lambda/4$: half-wave, quarter-wave plates, QP: thick quarts plates, BS: beamsplitter, PBS: polarization beamsplitters, $D_{i}$: single photon avalanche photodetectors, CC: coincidence circuit.}
\label{f:setup}
\end{center}
\end{figure}

To confirm the entanglement duality we use experimental setup depicted in Fig.~\ref{f:setup}. The setup consists of a Mach-Zehnder interferometer with non-polarizing beamsplitter (BS), number of half-wave and quarter-wave plates (HWP, QWP) and two polarizing beamsplitters (PBS) at the outputs of the interferometer, which share photons between distant users Alice and Bob. Such configuration is commonly used for polarization Bell-states measurements, implemented due to the presence of the symmetry of the momenta states. We, in turn, also demonstrate their entanglement, and show how the witness of a dualism can be probed.

A cw argon-ion laser is used to pump $2mm$ $\beta$-barium-borate (BBO) crystal. The crystal is cut for type-II phase matching and produces pairs of polarization-entangled photons with central wavelength $702.2nm$ propagated in two spatial modes $k$, $-k$ (selected by $3mm$ iris diaphragms, $1$m away from the crystal). The resulting walk-off effect arising due to crystal's birefringence is compensated by a combination of a half-wave plate at $45^\circ$ and a $1mm$ BBO crystal.
The initial setting of the setup is fixed on generating the output state $\left|\Psi\right\rangle=\tfrac1{\sqrt{2}}(|H\>_{k}|V\>_{-k}-|V\>_{k}|H\>_{-k})$. Polarization analysis provides a  visibility of interference curves $\approx97\%$ in the $H/V$ basis and $\approx96\%$ for the $+45/-45$ basis. This shows that we have produced a high-quality polarization-entangled photons source.

The photon pair is then sent to a Mach-Zehnder interferometer, formed by BBO crystal and BS, via the different input ports $-k$ and $k$. The relative path alignment $\Delta x$ to within the coherence length ($l\approx70\mu m$) is controlled by the optical trombone P1 with high precision.
After the BS photons are separated according to their polarization on PBSs, directed to Alice or Bob parts and detected on single photon avalanche photodiodes (SPADs) D1,D2,D3,D4 equipped with 5nm FWHM bandwidth interference filter and iris diaphragms.


For initial path length adjustment, the two-photon interference at BS is registered by monitoring the coincidence counts between detectors $D1,D4$ at orientation of the HWP at $45^\circ$ in one arm of the interferometer and varying $\Delta x$ by means of piezo drive.

\begin{figure}[ht]
\begin{center}
\includegraphics[width=0.5\textwidth]{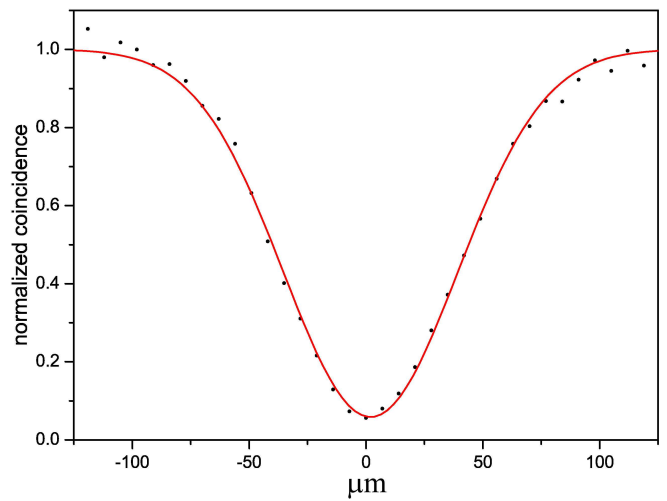}
\caption{Coincidence count rate as a function of the path length difference $\Delta x$. For perfect overlap destructive interference takes place, the observed visibility is $(90.4\pm0.4) \%$. The reduction of visibility can be attribute to non-equal reflection/transition indexes of BS (55/45) and non-perfect space overlapping. }
\label{f:HOM}
\end{center}
\end{figure}

Fig.\ref{f:HOM} shows a standard Hong-Ou-Mandel (HOM) anticoincidence dip for the overlap of two photons at a BS.

\section{Experimental Results}

For demonstration of the duality in entanglement we do two types of measurements with the same experimental setup (blocks a and b in  Fig.~\ref{f:setup}).
We start at first from the polarization correlation measurements on quantum state of two photons and verify their entanglement.
For this we investigate two-photon interference fringes by using HWPs at the front of the single photon detectors in  Fig.~\ref{f:setup} (block a). In the measurements of the two-photon interference fringes, we fix the HWP in one arm at 0 or $22.5^\circ$ and measure the coincidence count rates while rotating the orientation angle of HWP in the second arm. 
Clear two-photon interference fringes with high visibility were measured.


The next step is measuring the polarization entanglement-witness operator $W_{p}$. As we have mentioned before, the value of polarization entanglement-witness operator $W_{p}$ takes value $1$ for the maximum entangled state and falls to $0$ for mixed states. For pure non entangled states value  $W_{p}$ does not exceed 0.5. $W_p$ can be locally measured by choosing corresponding correlated measurements. For example, $|+45_{1}\>|+45_{2}\>\<+45_{1}|\<+45_{2}|$ is measured via correlations between detectors $D1,D4$, $|-45_{1}\>|-45_{2}\>\<-45_{2}|\<-45_{2}|$ between the pair $D2,D3$, at HWPs at $22.5^\circ$ in both arms. Operators $|R_{1}\>|L_{2}\>\<R_{1}|\<L_{2}|$ and $|L_{1}\>|R_{2}\>\<L_{1}|\<R_{2}|$ are measured at QWPs at $45^\circ$ in both arms by coincidence between detectors $D1,D3$ and $D2,D4$ correspondingly.

For the momentum entanglement, then the state can be written as (\ref{eq:state_mom}), and we essentially use the same expression of the entanglement-witness operator, just rewriting in the momentum degree of freedom, which is $W_m$ given by Eq.(\ref{eq:Wmom}). Consider two photons born in the SPDC process and propagating in two directions $-k$ and $k$. Four output ports from BS+PBSs are labelled by $k_{1},k_{2},k_{3},k_{4}$ and registered at the detectors $D1,D2,D3,D4$ (Fig.~\ref{f:setup} (block b)). Each detector registers whether a photon is in that port or not. It corresponds to measure the projection operators:

\begin{equation}
\begin{array}{cccc}
|k_{1}\>\<k_{1}|=\tfrac{(|k_{H}\>+i|-k_{H}\>)}{\sqrt{2}}\tfrac{(\<k_{H}|+i\<-k_{H}|)}{\sqrt{2}}\\

|k_{2}\>\<k_{2}|=\tfrac{(|k_{V}\>+i|-k_{V}\>)}{\sqrt{2}}\tfrac{(\<k_{V}|+i\<-k_{V}|)}{\sqrt{2}}\\

|k_{3}\>\<k_{3}|=\tfrac{(|k_{V}\>-i|-k_{V}\>)}{\sqrt{2}}\tfrac{(\<k_{V}|-i\<-k_{V}|)}{\sqrt{2}}\\

|k_{4}\>\<k_{4}|=\tfrac{(|k_{H}\>-i|-k_{H}\>)}{\sqrt{2}}\tfrac{(\<k_{H}|-i\<-k_{H}|)}{\sqrt{2}}\\
\end{array}
\end{equation}

Then momentum entanglement-witness operator $W_{m}$ is defined through the correlated measurements between four pairs of detectors: $D1,D2$; $D3,D4$; $D1,D3$; $D2,D4$.

\begin{equation}
\begin{array}{cc}
W_{m}=|k_{1}\>|k_{2}\>\<k_{1}|\<k_{2}|+|k_{3}\>|k_{4}\>\<k_{3}|\<k_{4}|- \\ (|k_{1}\>|k_{3}\>\<k_{1}|\<k_{3}|+
|k_{2}\>|k_{4}\>\<k_{2}|\<k_{4}|)
\label{eq:dualism}
\end{array}
\end{equation}

\begin{figure}
\begin{center}
\includegraphics[width=0.5\textwidth]{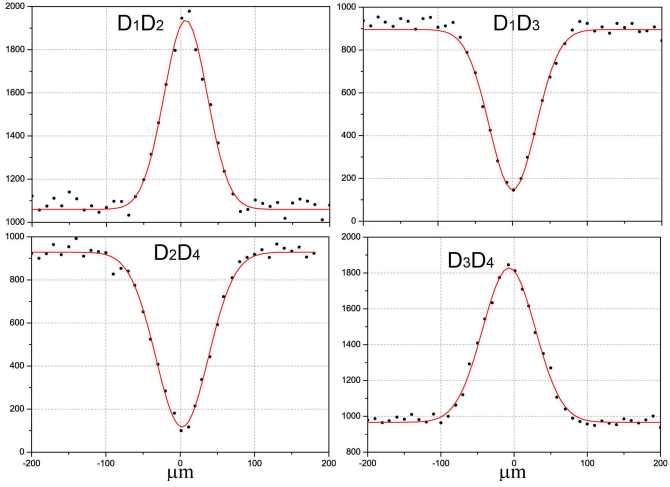}
\caption{Measurement of the coincidence rates by tuning the path length difference $\Delta x$ after readjusting the arm lengths of the interferometer.}
\label{f:4fig}
\end{center}

\end{figure}

\begin{figure}
\begin{center}
\includegraphics[width=0.5\textwidth]{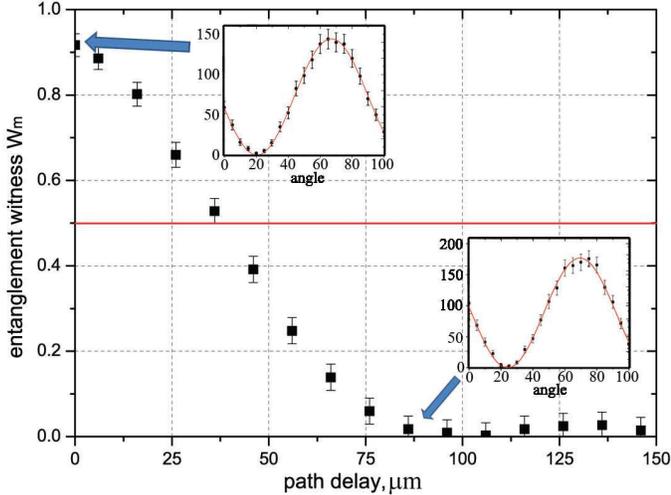}
\caption{Experimental results display the decrease of momentum entanglement witness $W_{m}$ as the path delay increases, while the inserted graphs show that polarization interference fringes remain unaffected for different values of $W_{m}$.}
\label{f:witness}
\end{center}
\end{figure}

\begin{table*}
\parbox{.45\linewidth}{
\centering
\begin{tabular}{|c|c|c|}
\hline
$\Delta x$, $\mu m$ & $W_m$ & $W_p$ \\
\hline
$0$ & $0.91\pm 0.03$ & $0.97 \pm 0.02$ \\
\hline
$46$ & $0.39\pm 0.03$ & $0.95\pm 0.02$ \\
\hline
$86$ & $0.02 \pm 0.03$ & $0.96\pm 0.02$ \\
\hline
\end{tabular}
\caption{\label{tab:wp}Experimental results of the witness measurement. First column is a path length difference between the two arms of the interferometer. Second and the third column are the entanglement witnesses for momentum and the polarization bases.}
}
\hfill
\parbox{.45\linewidth}{
\centering
\begin{tabular}{|c|c|c|}
\hline
$p$ & $W_p$ & $W_m$ \\
\hline
$0$ & $0.97\pm 0.02$ & $0.91\pm 0.03$ \\
\hline
$0.3$ & $0.33\pm 0.02$ & $0.87\pm 0.03$ \\
\hline
$1$ & $0.01 \pm 0.02$ & $0.87\pm 0.03$ \\
\hline
\end{tabular}
\caption{\label{tab:wp}Experimental results of witness measurement. First column specifies the degree of mixture introduced in the polarization degree of freedom. Second and the third column are the entanglement witnesses for momentum and the polarization bases. Duality in entanglement emerges only for indistinguishable photons}
}
\end{table*}
The value $W_{m}>0.5$ corresponds to an emergent momentum entangled state for the photon pair described by (\ref{eq:state_mom}). The data from the momentum entanglement measurements are shown in Fig. (\ref{f:4fig})

By moving path length different $\Delta x$ in the one arm of the interferometer we measure coincidences between four pairs of detectors: $D1,D2$; $D3,D4$; $D1,D3$; $D2,D4$, and the total number of events for different arriving time of photons on the BS. From these data we can reconstruct the momentum entanglement witness. For $\Delta x=0$ it takes maximum value $0.91\pm 0.03$, while for a nonzero delay the quantum state (\ref{eq:state_mom}) becomes mixed by momentum degree of freedom due to different arrival time on beamsplitter and the value of the momentum witness $W_{m}$ falls below 0.5 (Fig.\ref{f:witness}).
However, when we consider the polarization degree of freedom for the same state we observe a similarly large amount of the polarization witness $W_{p}$. Two inserted graphs in Fig.\ref{f:witness} show the sameness of the polarization interference fringes for different values of path delays corresponding to different degrees of mixedness for the momentum degree of freedom.
For a further exploration of the quantum feature of the duality we prepare also a mixed state of photons by polarization variable. The entanglement dualism has so far not been probed for a mixed state, which can arise because of decoherence.  To this end, we consider a density matrix of the polarization state:
\begin{equation}
\rho =(1-p)|\Psi_{pure}\rangle\langle\Psi_{pure}|+\frac{p}{2}(|H_{1}V_{2}\rangle\langle H_{1}V_{2}|+|V_{1} H_{2}\rangle\langle V_{1}H_{2}|)
\label{eq:family}
\end{equation}
where
\begin{equation}
| \Psi_{pure}\> = \tfrac1{\sqrt{2}}(|H\>_{1}|V\>_{2}-|V\>_{1}|H\>_{2})
\label{eq:Psi}
\end{equation}

is an initial pure maximal entangled polarization state and $p$ is a coefficient defining the degree of mixture.

In order to prepare a mixed polarization state and do not destroy the momentum entanglement it is necessary to introduce a quantum distinguishability in a controllable manner. We complement one of the output port of the interferometer by a set of quartz plates ( see Fig.\ref{f:setup}). A thick quartz plate with a vertically oriented optical axis introduced delay between vertically and horizontally polarized photons that led to their distinguishability, and hence, to the emergence of a mixture.

Thus, we destroy only polarization entanglement without affect on the arriving time of photons on the BS. It should be noted that there is no direct interaction on the BS between pairs of photons, because they have different polarizations (there is no direct two-photon interaction). That is why the thick quarts plate QP could be included in the "preparation" part of the setup, while PBSs, half-,quater- waveplates and detectors form the "measurement" part.

We have prepared three states of (\ref{eq:family}) with various degree of mixture: $p=1; 0.3; 0$ ($p=1$ corresponds to totaly mixed state, $p=0.3$ means that state contains $30\%$ of mixture). For each state we measure polarization and momentum entanglement witnesses. For $p=1$ we observed a flat interference fringe curve, for $p=0.3$ its visibility was reduced.
All collected data are presented in Tables I,II.

\section{Discussions}
We have verified a predicted dualism in the entangled state of two identical particles using the modification of a Bell-state apparatus.
Our apparatus is very different from the one in the theoretical proposal \cite{bose}, as well as the one reported in a recent test \cite{Duan}. Having an alternative method is always of benefit in the sense that one would probably intend to extend, in the future, such tests to entangled states of also material particles. In this sense they are sufficiently weakly interacting then the method proposed here is more advisable due to a less challenging phase stabilisation between different paths, with respect to the setup \cite{bose, Duan}. Furthermore, we have also verified a form of robustness of the states displaying the dualism -- even when decoherence or distinguishability deteriorates the entanglement of one of the two relevant degrees of freedom, the other may remain unaffected (maximally entangled). Thus, one has essentially two ways of exploiting the same resource state. This is best exemplified in Fig.\ref{f:setup}. If polarization entanglement is found to be persistently degraded by decoherence or an eavesdropper, then communicating parties Alice an Bob can use arrangement (b) to exploit the momentum entanglement in the state for quantum communications. This interesting property of the decoherence in one of the degrees of freedom not affecting the entanglement in terms of the other dual degree of freedom needs to be analyzed in depth in order to explore its further ramifications.

\section*{Acknowledgments}
This work was funded by NATO SPS Project 984397. E.V.Moreva acknowledges the support of Russian Foundation for Basic Research (project 13-02-01170-a). SB acknowledges support of the EPSRC Grant No. EP/J014664/1.

\end{document}